# Ferromagnetic Order Analysis of Fe- and FeO-modified Graphene-nano-ribbon

Norio Ota

Graduate School of Pure and Applied Sciences, University of Tsukuba, *1-1-1 Tenoudai Tsukuba-city 305-8571, Japan*

Graphene-nano-ribbon (GNR) is very attractive for ultra-high density spintronics devices. For checking a capability of self-induced ferromagnetic order, Fe- and FeO-modified GNR were analyzed based on the density functional theory. Model unit cells were [C32H2Fe2] and [C32H2Fe2O2]. Calculated results show the most stable spin state to be $Sz=8/2$ in [C32H2Fe2], whereas $Sz=6/2$ in [C32H2Fe2O2]. Magnetic moment M of Fe in [C32H2Fe2] was 3.65 $\mu_B$, which could be explained based on the Hund-rule considering donated charge to carbon to be $M^*=3.67 \mu_B$. There is a capability of ferromagnetic Fe spin array through an interaction with carbon $\pi$-conjugated spin system. There shows a long-range super-exchange order in [C32H2Fe2O2]. Optimized atomic configuration gave the typical 90 degree super-exchange coupling between Fe-3d and O-2p orbit. Magnetic moment of Fe by DFT was 2.58 $\mu_B$, whereas super-exchange model considering donated charge to oxygen gave an estimated magnetic moment to be 2.60 $\mu_B$. Sign of magnetic moment of Fe and O are all up-spin each other. We could expect ferromagnetic long-range-order as like a chain of ( –Fe-O-Fe-O-Fe- ). Band structure was analyzed in FeO-modified case, which suggested half-metal like behavior. We can expect several spintronics applications as like a spin filter.

## 1. Introduction

Current magnetic data storage[1)-3)] has a density around 1 tera-bit/inch$^2$ with 10 nm length, 25 nm width magnetic bit. We need a future ultra-high density spintronics material with a typical bit size of 1nm by 2.5nm. Recently, carbon based room-temperature ferromagnetic materials are experimentally reported[4)-9)]. They are graphite and graphene like materials. Kusakabe and Maruyama[10)-11)] theoretically proposed an asymmetric graphene-nano-ribbon (GNR) model with two hydrogen modified zigzag edge carbon showing ferromagnetic behavior. In our recent analysis[12-15)], several chemically modified graphene molecules and GNR also shows a possibility of strong magnetism. However, we need more large magnetization chemically modified GNR. Especially, we like to find a candidate with a self –induced ferromagnetic order one.

Here, by the first principles density functional theory (DFT) based analysis, Fe- and FeO-modified GNR were tried to find a ferromagnetic capability. Typical unit cell were [C32H2Fe2], and [C32H2Fe2O2].

Already, there were some experiments of graphene formation on Fe(110) substrate[16)], and Co layer structure observation on graphite substrate[17)]. In DFT calculation, iron cluster on graphene sheet[18)], or iron –based molecule grafted on graphene[19)] were studied. Those were the cases that iron atom coupled with graphene surface sheet. In a viewpoint of spintronics application, narrow stripe straight line like a GNR is suitable for industrial application. Therefore, this report focuses on chemically modified GNR.

## 2. Model graphene-ribbon

Bird eye view of Fe-modified GNR:[C32H2Fe2] is shown in Fig.2(a), whereas FeO-midified one:[C32H2Fe2O2] in (b). Modified Fe or FeO bonds with every zigzag edge carbon of one side (left hand side). Whereas, another side (right hand side) zigzag edges are all hydrogenated. Track width is 1.8 nm, tracking length of 1nm includes five zigzag carbon edges.

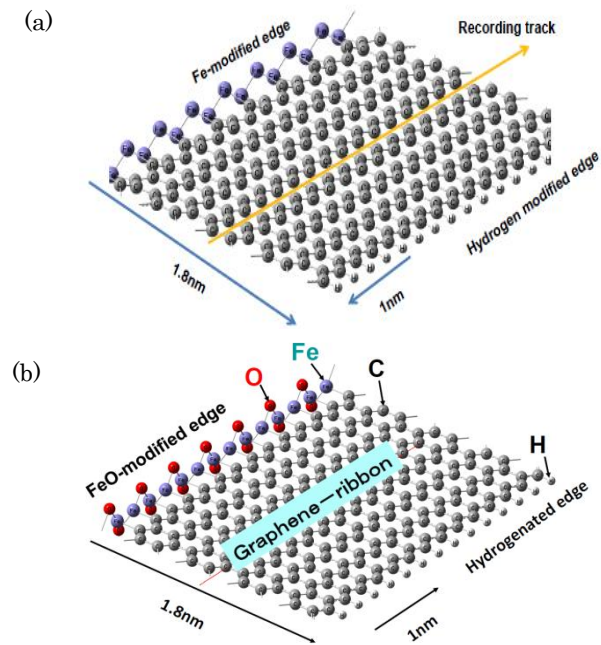

**Fig.1** Bird eye view of Fe-modified GNR (a) and FeO-modified one (b).

In Fig.2, an unit cell of Fe-modified GNR:[C32H2Fe2] is shown in a blue square mark, where blue ball show Fe, gray ball carbon and small ball hydrogen in a top view (a) and in a plane view (b). Whereas, unit cell of FeO-modified GNR:[C32H2Fe2O2] is demonstrated in Fig.3 as a plane view.

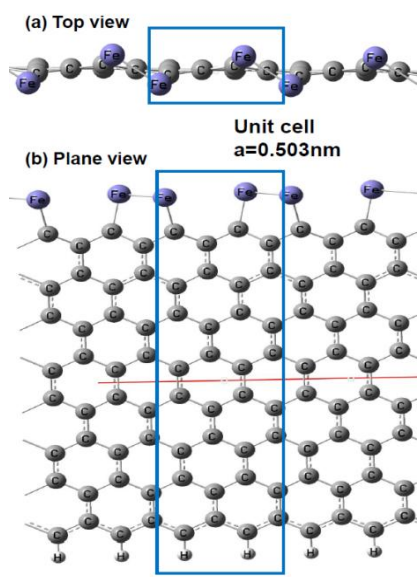

**Fig.2** Unit cell and DFT converged atomic configuration of Fe-modified GNR : [C32H2Fe2]. Blue ball shows Fe, gray one carbon and small one hydrogen in (a) top view and (b) plane view.

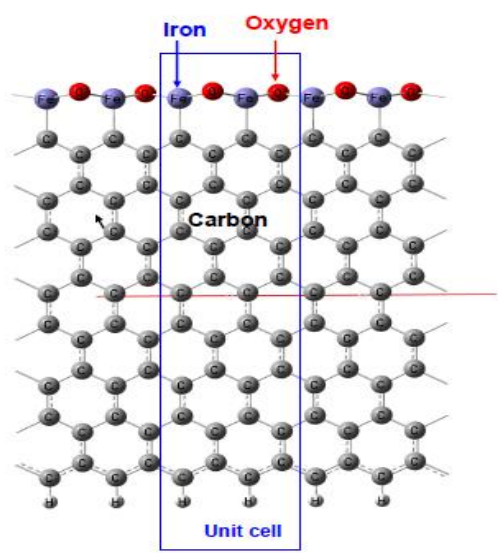

**Fig.3** Plane view of unit cell and DFT converged atomic configuration of FeO-modified GNR: [C32H2Fe2O2].

### 3. Calculation Method

We have to obtain the (1) optimized atom configuration, (2) total energy, (3) spin density configuration, (4) magnetic moment of every atom and (5) band characteristics depending on a respective given spin state Sz to clarify magnetism. Density functional theory (DFT) [20)-21)] based generalized gradient approximation (GGA-PBEPBE) [22)] was applied utilizing Gaussian03 package[23)] with an atomic orbital 6-31G basis set[24)]. In this paper, total charge of model unit-cell is set to be completely zero. Inside of a unit-cell, three dimensional DFT calculation was done. One dimensional periodic boundary condition was applied to realize an unlimited length GNR. Self-consistent calculations are repeated until to meet convergence criteria. The required convergence on the root mean square density matrix was less than 10E-8 within 128 cycles.

### 4, Fe-modified GNR [C32H2Fe2]

#### 4.1 Stable spin state and magnetic moment

Unit cell of Fe-modified GNR: [C32H2Fe2] is shown in Fig.2. Each unit cell has limited numbers of unpaired electrons, which enable six spin states like Sz=10/2, 8/2, 6/2, 4/2, 2/2 and 0/2. Starting DFT calculation, one certain Sz value should be installed as a spin parameter. Among them, DFT calculation resulted that the most stable spin state was Sz=8/2, which has 18.1 kcal/mol/unit-cell lower energy than that of Sz=10/2. In cases of spin states less than Sz=6/2, DFT calculations did not converged and suggested instability.

In case of Sz=8/2, optimized atomic configuration is shown in Fig.3 as a top view in (a) and a plane view in (b). Unit cell length "a" was 0.503nm. Every Fe atom positioned a little bit out of plane from major GNR carbon part. We could find a detailed structure of Fe-modified part in Fig.4. Distance between Fe-C (zigzag edge) was 0.191nm and Fe-Fe 0.218nm, whereas Fe to another side Fe distance was 0.345nm. Tilt angle of Fe from GNR plane was +/- 14 degree.

Spin density of Sz=8/2 was illustrated in Fig.5. Red cloud shows up-spin density, whereas blue cloud down-spin one. Contour lines for 0.001, 0.1, 0.1e/A$^3$ are predicted by arrows. Iron atom wears very large up-spin cloud. Bonded zigzag edge carbon shows down-spin, whereas next neighbor carbon up-spin. Inside of GNR, up and down spins alternately arranges one by one very regularly [15)].

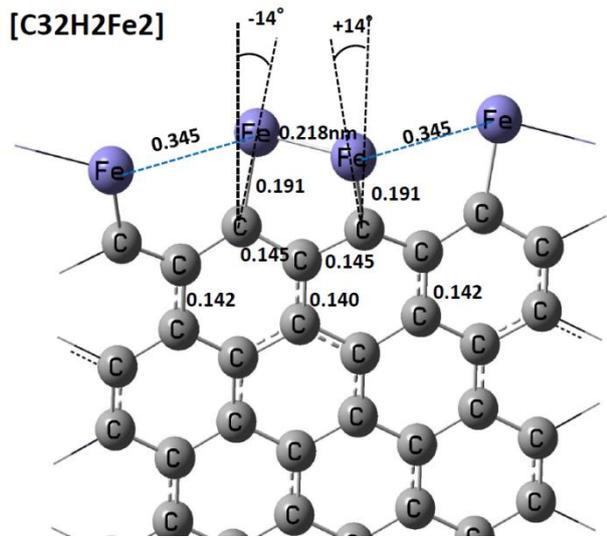

**Fig.4** Converged atomic configuration of [C32H2Fe2] focused on Fe-modified part of GNR. Position of all Fe were out of plane from carbon part of GNR.

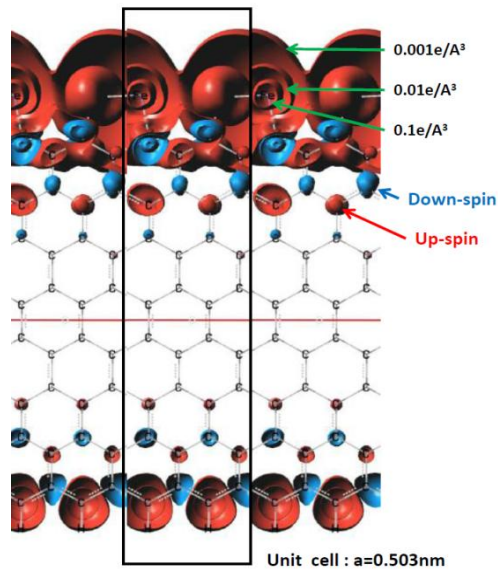

**Fig.5** Spin density configuration in [C32H2Fe2]. Red cloud shows up-spin, whereas blue one down-spin.

DFT calculation gives an atomic magnetization M. In case of Sz=8/2, single iron:Fe has M(Fe)=3.65 $\mu_B$. We tried to explain this value based on the Hund-rule[25]. As illustrated in Fig.6, single isolated Fe has 4 $\mu_B$. We should subtract escaped charge 0.33e from Fe-3d up-spin site. By such a simple assumption, we could estimate magnetic moment M*(Fe) to be 4.00-0.33=3.67 $\mu_B$. Those M and M* values are very close together,

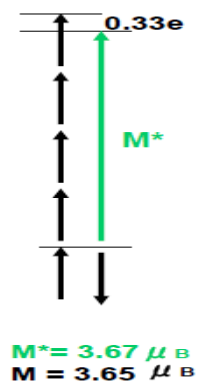

**Fig.6** Hund-rule based iron:Fe magnetization is estimated to be M*(Fe)=3.67, whereas DFT calculated one M(Fe)=3.65 $\mu_B$.

### 4-2, Capability of ferromagnetic Fe spin array

Magnetization of Fe was simply explained by a modified Hund-rule. This means that Fe-atom plays as an almost isolated atom. Also, DFT calculation predicted that Fe atom did not exchange-coupled directly with a neighbor Fe each other. However, it should be noted that, in a graphene system, short range order between Fe and C causes remarkable magnetic order in a whole system.

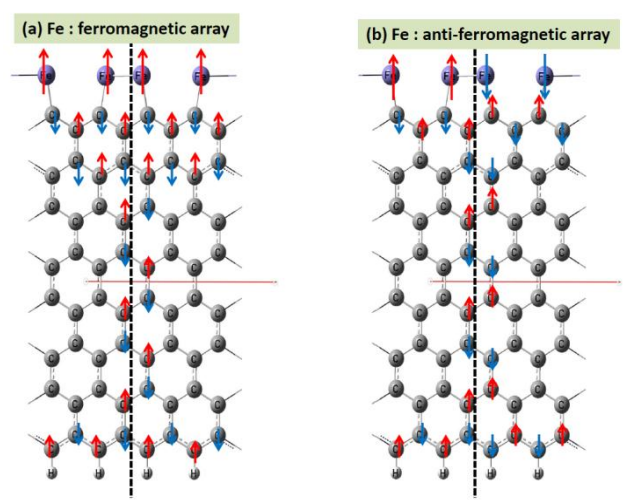

**Fig.7** Ferromagnetic Fe-spin array and carbon site spin configuration in (a), whereas anti-ferromagnetic case in (b). In (b), at a unit cell boundary (dotted line), there causes irregular up-up and down-down spin pairs. Stable spin configuration is (a).

We tried a spin arrangement consideration in doubly enhanced unit cells. All Fe up-spin case as shown in Fig.7 (a), there is a reasonable spin arrangement inside of graphene, that is, up-down spin-pairs are align vey regularly at whole carbon sites of GNR. Such spin configuration decrease an exchange energy and finally stabilize total system. On the contrary, in case of two up-spin and two down-spin case in (b), there causes many up-up and down-down spin-pairs on a boundary of unit cells shown by a dotted line. Such spin configuration irregularly elevates total energy and causes instability. It looks that there is a ferromagnetic like Fe spin array by an interaction between Fe and carbon $\pi$-conjugated spin system. All of Fe-spins will become up-spin and may give macroscopic ferromagnetic feature.

### 5, FeO-modified GNR [C32H2Fe2O2]

#### 5.1 Stable spin state and atomic configuration

Capable spin state of [C32H2Fe2O2] are Sz=8/2, 6/2, 4/2, 2/2 and 0/2. Among them, the most stable spin state was Sz=6/2. Spin state Sz=8/2 show 43kcal/mol/unit-cell larger energy than Sz=6/2. In cases of Sz less than 4/2 were all unstable.

In case of Sz=6/2, optimized atomic configuration is shown in Fig.8 as a bird eye view focused on FeO-modified part. It should be noted that the optimized angle between Fe-O-Fe was +91 degree, and next Fe-O-Fe was -91 degree. Distance between Fe-O was 0.177nm, whereas distance Fe-C was 0.183nm.

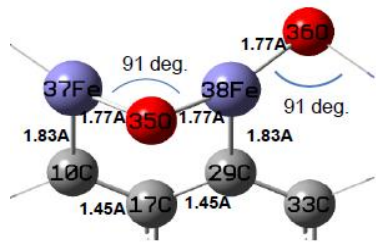

**Fig.8** A bird eye view of FeO-modified GNR:[C32H2Fe2O2] focused on a FeO-modified top pat. Angle of Fe-O-Fe was +/- 91 degree.

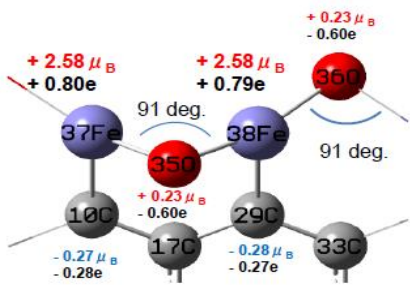

**Fig.9** Atomic magnetic moment and charge of [C32H2Fe2O2] overlapped on a bird eye view of FeO-modified part.

#### 5.2 Magnetic moment and super-exchange coupling

Atomic magnetic moment and charge of FeO-modified part were illustarated in Fig.9. Fe-atom has a DFT converged magnetic moment of M(Fe)=+2.58 $\mu_B$. Whereas, zigzag edge carbon has a minus sign of M(C)=-0.27 $\mu_B$. It should be noted that oxygen also has a magnetic moment of +0.23 $\mu_B$. Charge of Fe was +0.80e, which means that Fe donated extra charge to the nearest carbon (-0.28e) and oxygen (-0.60e).

We could see beautiful 90 degree Fe-O arrangement. Here, we like to explain those magnetic moments by 90 degree super-exchange model bringing self-induced magnetic-order.

In Fig.10, typical 90 degree Fe-O-Fe super-exchange model is illustrated. Electrons of Fe-$3d^6$ are exchange-coupled with O-$2p^4$ orbital electrons. A part of those $2p^4$ electrons are donated by Fe itself and excited to be down-spin as shown by a blue dotted arrow. This excited down-spin had exchange coupled with Fe-$3d^6$ electrons and stabilized total spin configuration. There happens similar spin configuration on a right hand side coupling between O and Fe. By such mechanism, there causes Fe(up-spin)-Fe(up-spin) arrangement. It should be noted that there is a capability of macroscopic self-induced ferromagnetic order by such an array of (-Fe-O-Fe-O-).

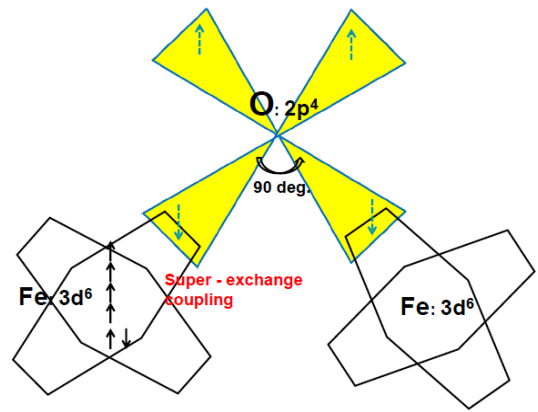

**Fig.10** Simple illustration of 90 degree super-exchange model between Fe-$3d^6$ and O-$2p^4$ orbits.

Based on the simple Hund-rule, magnetic moment of Fe =M*(Fe) was estimated as shown in Fig.11. Escaped charge 0.80e from Fe (calculated by DFT) should be subtracted from Fe up-spin site. This escaped charge partially donated to oxygen 0.60e, and the rest 0.20e to neighbor carbons. This 0.60e charge was excited as oxygen-2p orbits and super-exchange-coupled with Fe as down-spin configuration as shown by broken-arrow to be 0.60 $\mu_B$. Total M*(Fe) is estimated to be 2.60 $\mu_B$

(=4.00-0.80-0.60). This M(Fe) value is very close with the DFT calculated result M(Fe)=2.58 $\mu_B$.

Those results brings a capability of self-induced long-range magnetic order, that is, ferromagnetic spin array on GNR as like typically imaged in Fig.12 and Fig.14. Fig12 shows a spin density map viewed from a top side. We can see a very high density (0.1e/A$^3$) up-spin cloud around Fe-atom, also can see a up-spin bridge structure (0.01e/A$^3$) at oxygen site combining Fe to Fe.

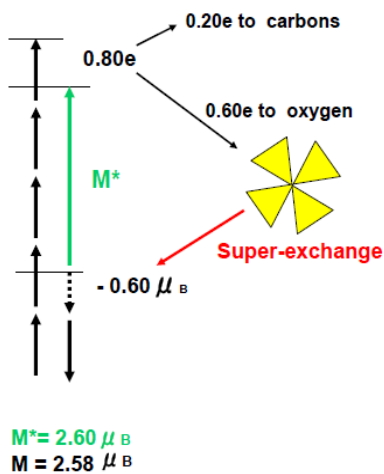

**Fig.11** Estimation of Hund-rule based Fe magnetic moment M*(Fe) considering Fe-O super-exchange coupling.

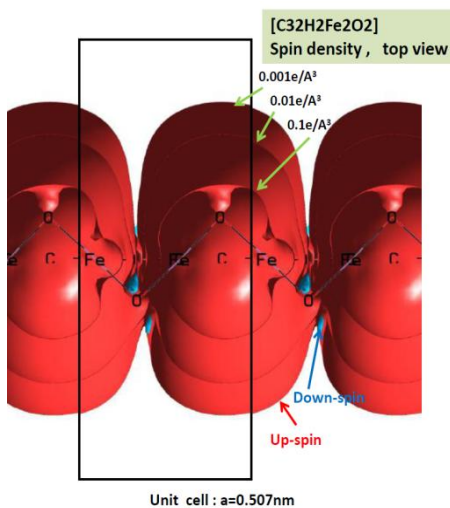

**Fig.12** Spin density top view of [C32H2Fe2O2]. There appears up-spin bridge at oxygen position, which suggest a super-exchange coupling between Fe-O-Fe.

### 5.3 Fe-carbon π-conjugated system

In Fig.13, we can see a total spin density configuration of [C32H2Fe2O2] obtained by DFT calculation. There is a very large up-spin cloud at FeO-modified part. Whereas, there appear up- and down regularly aligned spin arrangement at carbon sites. This origin is the same with Fe-modified GNR, that is, Fe-carbon π-conjugated spin arrangement. Fig.14 is a total spin arrangement showing two origins of Fe-Fe ferromagnetic array. One is the super-exchange on (-Fe-O-Fe-) system, and another Fe-C π-conjugated system.

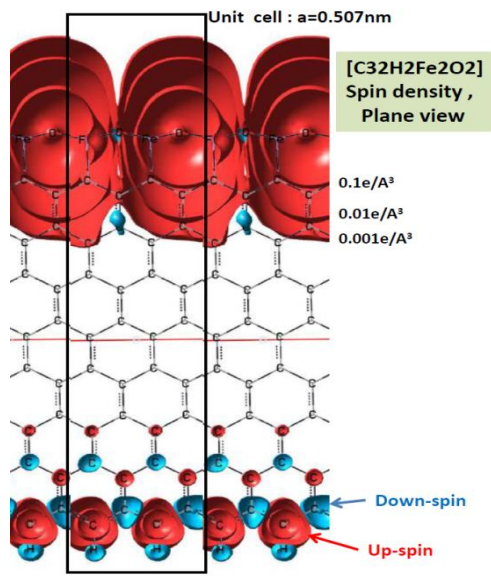

**Fig.13** Spin density configuration of [C32H2Fe2O2]. There is a very large up-spin cloud at FeO-modified part. Whereas, there appear up- and down-spin carbon π-conjugated spin arrangement inside of GNR

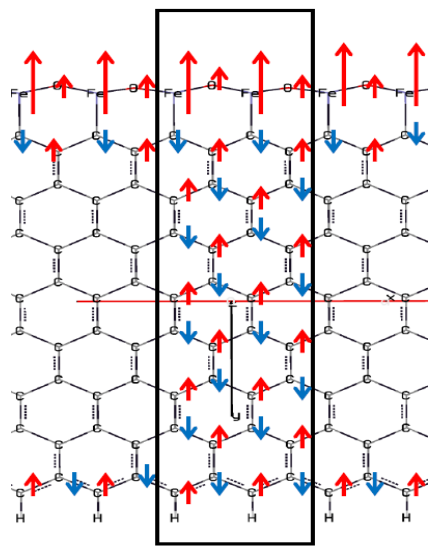

**Fig.14** Total spin arrangement of [C32H2Fe2O2]. There appear -Fe-O-Fe- long range order and also π-conjugated carbon spin order.

## 5.4. Band characteristics of FeO-modified GNR

GNR is a periodic system with one-dimensional crystallography. Here, Band characteristics of [C32H2Fe2O2] were analyzed as shown in Fig.15. Lattice parameter "a" is 0.507nm. We divided k-space to 12 elements from k=0/a to $\pi$/a. In Fig.15, red curves are up-spin occupied orbits, light red unoccupied one, blue down-spin occupied and light blue unoccupied.

Up-spin energy gap was 0.65eV. Among this gap, there were several unoccupied down-spin orbits. Density of state, as shown in a center part of Fig.15, suggested a down-spin dominated half metal like characteristics. In case of half metal, we can expect several spintronics devices as like a spin filter.

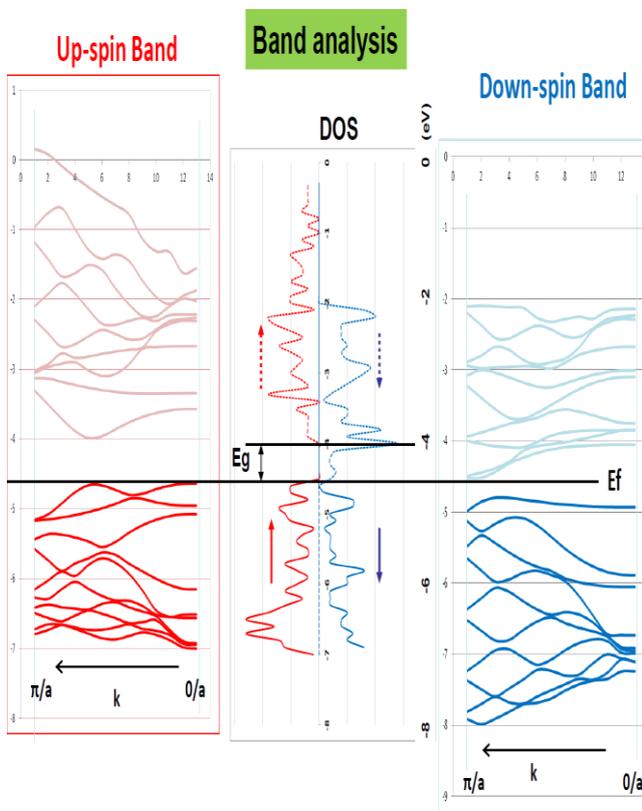

**Fig.15** Band characteristic of [C32H2Fe2O2].
Up-spin energy gap is 0.55eV. Among this gap, there are several down-spin orbits.

## 8. Conclusion

Graphene-nano-ribbon is very attractive for an ultra-high density spintronics devices. In order to check a capability of self induced ferromagnetic order, Fe- and FeO-modified GNR were analyzed based on the density functional theory. Model unit cells were [C32H2Fe2] and [C32H2Fe2O2].

(1) DFT calculation show the most stable spin state to be Sz=8/2 in [C32H2Fe2], whereas in [C32H2Fe2O2], Sz=6/2.
(2) Magnetic moment M of Fe in [C32H2Fe2] obtained by DFT was 3.65 $\mu_B$, which was explained based on the Hund-rule considering donated charge +0.33e to carbon, which gave a magnetic moment M* to be 3.67 $\mu_B$. There is a capability of ferromagnetic like Fe spin array through an interaction with carbon $\pi$-conjugated system.
(3) There shows a long-range super-exchange order in [C32H2Fe2O2]. Optimized atomic configuration gives typical 90 degree super-exchange coupling between Fe-3d and O-2p orbit. Magnetic moment of Fe by DFT was 2.58 $\mu_B$, whereas Hund-rule based super-exchange model considering donated charge to Oxygen gave an estimated magnetic moment to be 2.60 $\mu_B$. Sign of magnetic moment of Fe and O are all up-spin. We could expect ferromagnetic long-range-order as (–Fe-O-Fe-O-Fe-).
(4) Band structure was analyzed in FeO-modified case. Down-spin dominated half-metal like behavior was obtained. We can expect several applications for spintronics devices as like a spin filter.